\documentclass[useAMS]{mnras}
\usepackage{amsmath}
\usepackage{gensymb}
\usepackage{textcomp}
\usepackage{amssymb}
\usepackage[pdftex]{graphicx}
\usepackage{amssymb}
\usepackage{esint}
\usepackage{physics}
\usepackage{stfloats}
\usepackage[section]{placeins}
\usepackage[toc,page]{appendix}
\usepackage{multirow}
\usepackage{hyperref}
\usepackage{soul}
\usepackage[bottom]{footmisc}
\usepackage[notquote]{hanging}

\begin{document}
\raggedbottom
\definecolor{britishracinggreen}{rgb}{0.0, 0.26, 0.15}
\title{Line Driven Winds from Variable Accretion Discs}
\author[A. Kirilov, S. Dyda, C. S. Reynolds]
{\noindent\parbox{\textwidth}{Anthony Kirilov\thanks{antoniikirilov@gmail.com}, Sergei~Dyda\thanks{sdyda@ast.cam.ac.uk}, Christopher S. Reynolds}\\
Institute of Astronomy, Madingley Road, Cambridge CB3 0HA, UK
}

\date{\today}
\pagerange{\pageref{firstpage}--\pageref{lastpage}}
\pubyear{2022}
\setlength\parindent{24pt}

\label{firstpage}

\maketitle

\begin{abstract}
\color{black}
We use numerical hydrodynamics simulations to study line driven winds launched from an accreting $\alpha-$disc. Building on previous work where the driving radiation field is static, we compute a time-dependent radiation flux from the local, variable accretion rate of the disc. We find that prior to the establishment of a steady state in the disc, variations of $\sim 15\%$ in disc luminosity correlate with variations of $\sim 2-3$ in the mass flux of the wind. After a steady state is reached, when luminosity variations drop to $\sim 3\%$, these correlations vanish as the variability in the mass flux is dominated by the intrinsic variability of the winds. This is especially evident in lower luminosity runs where intrinsic variability is higher due to a greater prevalence of failed winds. The changing mass flux occurs primarily due to the formation of clumps and voids near the disc atmosphere that propagate out into the low velocity part of the flow, a process that can be influenced by \emph{local} variations in disc intensity. By computing the normalised standard deviation of the mass outflow, we show that the impact of luminosity variations on mass outflow is more visible at higher luminosity. However, the absolute change in mass outflow due to luminosity increases is larger for lower luminosity models due to the luminosity-mass flux scaling relation becoming steeper. We further discuss implications for CVs and AGN and observational prospects.
\end{abstract}


\section{Introduction}
\label{sec:introduction}
Many accretion disc systems such as cataclysmic variables (CVs), X-ray binaries (XRBs) and active galactic nuclei (AGN) exhibit blue shifted absorption lines, which is interpreted as evidence of outflowing gas. In general, outflows may be accelerated via thermal, radiation or magnetic pressure. An important question for any system exhibiting outflows is which of the above mechanisms, or combination thereof, acts to launch and accelerate this gas? 

In the case of CVs and AGN a strong candidate acceleration mechanism is radiation pressure on spectral lines, so called line driving. Observationally this is particularly interesting because line driving directly couples accretion and outflow energy. Matter accreting in the disc converts gravitational potential energy into photon energy. These photons irradiate the gas above the disc and drive an outflow by transferring momentum from the radiation field to the gas. Drawing intuition from spherically symmetric models of line driven winds (Castor, Abbot \& Klein 1975, hereafter CAK75), for isothermal winds with a fixed chemical abundance, the strength of the outflow (mass flux and outflow velocity) should depend strongly on the driving luminosity and not so much on the density at the base of the outflow. This is in contrast to thermally driven winds (Parker 1958), where for an isothermal wind, the mass flux has a steep dependence on density at the base.

Though only approximately correct in non-spherical geometries, the correlation between accretion energy and outflow properties for line driven winds can be exploited to constrain relevant physical processes. In many systems we observe both the total system luminosity and discern properties of the outflowing gas by measuring spectra and using photoionization modeling. Self-consistent treatment of such a system therefore requires simulating both the physics in the accretion disc and the outflow. 

Originally, line driving, operating  in  the  Sobolev  approximation,  was  proposed  as  a driving mechanism for stellar winds in OB stars (Lucy \& Solomon 1970, hereafter LS70 \& CAK75) and was successful in predicting their mass loss rates and outflow velocities (Pauldrach, Puls \& Kudritzki 1986, Friend \& Abbott 1986). Line driving has since been applied to a variety of systems, in particular those with accretion discs such as CVs and AGN.

Early work studied line driven disc winds using a variety of simplifying assumptions. Early analytical studies assumed simple scaling for the
strength  of  the  radiation  field  (Vitello  \&  Shlosman  1988) or a decoupling of the radial and polar angle equations of motion (Murray et. al 1995). Later simulations found stationary outflow solutions (Pereyra,  Kallman \&  Blondin  (1997); Pereyra (1997)), but these were too coarse to properly resolve the critical point. This transition was resolved using non-uniform meshes near the disc midplane (Proga, Stone \& Drew 1998; 1999) which found that as in the CAK picture the strength of the outflow was correlated with total system luminosity but the non-spherical geometry led to these flows being unsteady for low driving luminosity. Later work in 3D showed that relaxing the axisymmetry assumption leads to the formation of clumps (Dyda \& Proga 2018a,b hereafter DP18a,b). The geometry of the radiation field can also affect the strength of the outflow by altering the flow geometry (Dyda \& Proga 2018c). 

In this prior work the disc served as a matter reservoir and winds were driven by a \emph{time-independent} radiation field. Some later studies sought to loosen this assumption on the radiation field. Kurosowa \& Proga (2009) used the variable mass at the inner boundary to estimate the accretion rate and corresponding disc luminosity and found strong correlations between these and the wind outflow. These models were extended to include reprocessed and scattered photons, which were found to further enhance the strength of the outflow (Liu et al. 2013; Mosallanezhad et al. 2019). Nomura et al. (2020) computed disc luminosity by accounting for mass losses from the wind and found these losses alter the disc UV continuum responsible for much of the line driving flux.

We build on prior models by allowing for a \emph{time-dependent} radiation field, computed self consistently from the accretion in the disc. We simulate a thin disc where viscous dissipation is computed using an $\alpha$-prescription (Shakura \& Sunyaev 1973) and the local disc intensity is computed from the local accretion rate. The luminous accretion disc drives an outflow via radiation pressure on spectral lines. We carry out a series of simulations for different time-averaged disc luminosities and characterize the properties of the resulting outflows and their correlation with disc properties.

The structure of our paper is as follows. In Section \ref{sec:numerical} we describe our numerical methods, including how the disc intensity is computed from the accretion rate. In Section \ref{sec:results} we describe the results of varying accretion rate on the time-averaged and time-dependent outflow properties. We conclude in Section \ref{sec:discussion} where we discuss the implications for CVs and AGN winds and observational prospects.

\section{Numerical Methods}
\label{sec:numerical}
We performed all numerical simulations with the publicly available MHD code \textsc{Athena++} (Stone et al. 2020). The basic physical setup is a gravitating, central object surrounded by a thin, axisymmetric, luminous $\alpha$-disc. The accretion disc acts as a source of driving radiation, accelerating the gas that is assumed to be optically thin to the continuum.  Our radiation model is described in detail in DP18a,b, but we sketch the basic setup here.

\subsection{Basic Equations}
\label{sec:basic_equations}
The basic equations for single fluid hydrodynamics driven by a radiation field are
\begin{subequations}
\begin{equation}
\frac{\partial \rho}{\partial t} + \nabla \cdot \left( \rho \mathbf{v} \right) = 0,
\end{equation}
\begin{equation}
\frac{\partial (\rho \mathbf{v})}{\partial t} + \nabla \cdot \Big[\rho \mathbf{vv} + \mathbf{P} + \mathsf{\tau} \Big] = - \rho \nabla \Phi + \rho \mathbf{F}^{\rm{rad}},
\label{eq:momentum}
\end{equation}
\begin{equation}
\frac{\partial E}{\partial t} + \nabla \cdot \Big[ (E + P)\mathbf{v} + \left(\tau \cdot \mathbf{v}\right) \Big] = -\rho \mathbf{v} \cdot \nabla \Phi + \rho \mathbf{v} \cdot \mathbf{F}^{\rm{rad}} ,
\label{eq:energy}
\end{equation}
\label{eq:hydro}%
\end{subequations}
where $\rho$, $\mathbf{v}$ are the fluid density and velocity respectively, $\mathbf{P}$ is a diagonal tensor with components P the gas pressure, $\mathsf{\tau}$ is the viscosity tensor and $\mathbf{F}^{\rm{rad}}$ is the radiation force. For the gravitational potential, we use $\Phi = -GM/r$ and $E = 1/2 \rho |\mathbf{v}|^2 + \mathcal{E}$ is the total energy where $\mathcal{E} =  P/(\gamma -1)$ is the internal energy. The isothermal sound speed is $a^2 = P/\rho$ and the adiabatic sound speed $c_s^2 = \gamma a^2$. We use a nearly isothermal equation of state $P = k \rho^{\gamma}$ with $\gamma = 1.01$.   The temperature is then $T = (\gamma -1)\mathcal{E}\mu m_{\rm{p}}/\rho k_{\rm{b}}$ where $\mu = 0.6$ is the mean molecular weight and other symbols have their standard meaning.

We model the viscosity using the Shakura-Sunyaev $\alpha-$disc prescription, where the kinematic viscosity is given by
\begin{equation}
\nu = \alpha_{\nu} \frac{c_s^2}{\Omega_K},
\end{equation}
for dimensionless parameter $ \alpha_{\nu} = 0.01$ and $\Omega_K = \sqrt{GM/r^3}$ the Keplerian orbital frequency.

\subsection{Radiation Force and Accretion Disc}
\label{sec:radiation_force}
We assume a time-dependent radiation field, computed from the local accretion rate of the axisymmetric, thin disc along the midplane. The frequency integrated intensity of a thin disc is
\begin{align}
I(r_d) &= \frac{3}{\pi} \frac{GM}{r_{\star}^2}\frac{c}{\sigma_e} \Gamma_d \left( \frac{r_{\star}}{r_d} \right)^3 \left[ 1 - \left(\frac{r_{\star}}{r_d} \right)^{1/2}\right] ,
\label{eq:intensity}
\end{align}   
where
\begin{equation}
\Gamma_d = \frac{\dot{M}_{\rm{acc}} \sigma_e}{8 \pi c r_{\star}}
\end{equation}
is the disc Eddington number, $\dot{M}_{\rm{acc}}$ is the accretion rate in the disc (see for example Pringle 1981), $\sigma_e$ is the Thompson cross section, $r_d$ is the radial position on the disc, and $r_{\star}$ the inner radius of the disc and $c$ the speed of light. The intensity profile (\ref{eq:intensity}) assumes a time independent accretion rate throughout the disc and therefore a constant Eddington fraction.  

In DP18a,b we assumed a fixed $\dot{M}_{\rm{acc}}$ but now we compute it within the simulation domain. We break up the disc into rings and compute a local accretion rate
\begin{equation}
    \dot{M}_{\rm{acc}}(r,t) = 4 \pi \int_{\theta_d}^{\pi/2} \rho v_{r} r^2 \sin \theta d \theta, 
\label{eq:local_acc}
\end{equation}
where $\theta_d$ is at the surface of the disc as defined by a density floor $\rho_{d}=10^{-10}$. We substitute this accretion rate into (\ref{eq:intensity}) to compute the local disc intensity. This is an approximation, since we use the analytic expression for the local disc intensity that was derived assuming a constant accretion rate for the global disc, not just a local patch. An alternative approach would be to compute the local intensity from the viscous dissipation. We chose our approach to simplify the comparison between coupled and uncoupled models. The radiation force is then computed by assuming the gas is optically thin to this radiation field and every point in the wind experiences a radiation force
\begin{equation}
\mathbf{F}^{\rm{rad}} = \mathbf{F}^{\rm{rad}}_e + \mathbf{F}^{\rm{rad}}_{L},
\end{equation}  
which is a sum of the contributions due to electron scattering $\mathbf{F}^{\rm{rad}}_e$ and line driving $\mathbf{F}^{\rm{rad}}_{L}$. In this continuum, optically thin approximation, the radiation force due to electron scattering is
\begin{equation}
\mathbf{F}^{\rm{rad}}_e = \varoiint \left( \mathbf{n} \frac{\sigma_e I d\Omega}{c} \right),
\end{equation} 
where $\sigma_e$ is the electron scattering cross section, $\mathbf{n}$ is the normal vector from the radiating surface to the point in the wind, $d\Omega$ is the solid angle and the integration is carried out over the entire disc. 

We treat the radiation due to lines using a modification of the CAK formulation where the radiation force due to lines is 
\begin{equation}
\mathbf{F}^{\rm{rad}}_L = \varoiint {\cal{M}} (t) \left( \mathbf{n} \frac{\sigma_e I d\Omega}{c} \right),
\label{eq:f_line}
\end{equation}
and ${\cal{M}} (t)$ is the so-called force multiplier. We use the Owocki, Castor \& Rybicki (1984) parametrization of the line strength, where working in the Sobolev approximation,
\begin{equation}
{\cal{M}} (t) = k t^{-\alpha} \left[ \frac{(1 + \tau_{\rm{max}})^{1-\alpha} - 1}{\tau_{\rm{max}}^{1 - \alpha}}\right],
\label{eq:force_multiplier}
\end{equation} 
where $k$ and $\alpha$ are constants, $\tau_{\rm{max}} = t \eta_{\rm{max}}$, $\eta_{\rm{max}}$ is related to the maximum force multiplier via ${\cal{M}}_{\rm{max}} = k (1 - \alpha) \eta_{\rm{max}}^{\alpha}$ and the optical depth parameter
\begin{equation}
t = \frac{\sigma_e \rho v_{\rm{th}}}{|dv_{l}/dl|},
\label{eq:optical_depth_parameter}
\end{equation} 
where $v_{\rm{th}}$ is the thermal velocity of the gas and $dv_{l}/dl$ is the velocity gradient along the line of sight. We take $k = 0.2$, $\alpha = 0.6$ and $M_{\rm{max}} = 4 400$. Additional details about our numerical treatment of the radiation force can be found in the Appendix of DP18a and DP18b.

\subsection{Simulation Setup}
\label{sec:simulation_parameters}

The initial conditions consists of a disc in hydrostatic equilibrium with density profile
\begin{equation}
\rho = \rho_0 \left(\frac{r_{\star}}{r_d}\right)^3 \left(1-\sqrt{\frac{r_{\star}}{r_d}}\right) \exp(-z^2/2h^2),
\label{eq:density}
\end{equation}
with $r_d$ the radial position on the disc,  and the scale-height $h = c_s / \Omega_K$. The midplane density parameter $8.33 \times 10^{-3} \leq \rho_0 \leq 1.6 \times 10^{-1}$ (in $\rm{g}\ \rm{cm}^{-3}$) is chosen so the accretion rate at late times produces the same disc luminosity as our uncoupled disc runs. For reference, the lowest luminosity run has accretion rate of $\sim 9.8 \times 10^{-9} M_{\odot}\ \rm{yr}^{-1}$ at late times. The angular velocity is initialized to its Keplerian value, $\Omega_K=\sqrt{GM/r_d^3}$.

The  radial  computational  domain  extends  over  the  range $r_{\star} \leq r \leq 16 r_{\star}$ with  logarithmic  spacing  between grids $\Delta r_{i+1}/\Delta r_{i}=1.05.$  The  polar  angle  range  is  $0 \leq \theta \leq \pi/2$  and  has  logarithmic  spacing $\Delta \theta_{i+1}/\Delta \theta_{i}=0.95$ that  ensures  that  we  have  sufficient  resolution near the disc midplane to resolve disc accretion and wind acceleration. We use a grid resolution of $n_{r}$×$n_{\theta}$=96×96 cells and $N_{r}$×$N_{\theta}$=3×3=9 MPI meshblocks.

We impose outflow boundary conditions at the inner and outer radial boundaries, axis boundary conditions along the $\theta = 0$ axis and reflecting conditions about the  $\theta = \pi/2$ midplane.

We chose parameters characteristic of a CV system. The  central  object  has  mass  and  radius M=0.6 $M_{\odot}$ and $r_{\star} = 8.7 \times 10^8$ cm, respectively. The orbital period at the inner radius is then $T_0=18$ s. The ratio between the gas thermal energy and gravitational potential energy, sometimes referred to as the hydrodynamic escape parameter,  $\rm{HEP} = GM/r_{\star} c_{s}^2   =8 \times 10^3$ at the base of the wind, corresponding to a sound speed $c_s \approx 3.4 \times 10^{6}$ cm\ $\rm{s}^{-1}$. For this value of HEP, thermal driving, which requires $\rm{HEP} \lesssim 10$ (Stone \& Proga 2009; Dyda et al. 2017), is negligible throughout the domain but it is not too high that we cannot resolve the disc accretion with our resolution. The density throughout the domain outside the disc is set to $\rho=10^{-20}\ \rm{g}\ \rm{cm}^{-3}$. We impose a density floor of $10^{-22}\ \rm{g}\ \rm{cm}^{-3}$ and a pressure floor computed via our nearly isothermal, adiabatic equation of state.

\section{Results}
\label{sec:results}
\setlength\parindent{24pt}

\begin{table*}
  
  \begin{center}
    \begin{tabular}{c |c |c | c | c | c | c | c | c |}
      Model & Type & $\rho_0\ [\rm{g}\ \rm{cm}^{-3}]$ & $L_{\rm{avg}}{\cal{M}}_{\rm{max}}$ [$L_{Edd}$]& $\dot{M}_{\rm{avg}}$[$M_{\odot}\ \rm{yr}^{-1}$] & $\sigma_{ L}/   L_{\rm{avg}}$ & $\sigma_{\dot{M}}/\dot{M}_{\rm{avg}}$ & $\Delta \dot{L}_{\rm{max}} / \dot{L}_{\rm{avg}}$ & $\Delta \dot{M}_{\rm{max}}/\dot{M}_{\rm{avg}}$ \\ \hline
U1  & Uncoupled & $8.33 \times 10^{-3}$ & $1.7$ & $2.05 \times 10^{-13}$&$0.06$ &$0.35$ & $0\%$ & $ 200\%$\\
C1  & Coupled & $8.33 \times 10^{-3}$ & $1.7$ &  $3.75 \times 10^{-13}$ & $0.06$ &   $0.36$ & $ 15\%$ & $200\%$\\
\\
U2  & Uncoupled & $2.00 \times 10^{-2}$& $4.1$& $3.87 \times 10^{-12}$ &$0.06$ &  $0.18$ & $0\%$ & $ 50\%$\\
C2  & Coupled & $2.00 \times 10^{-2}$& $4.1$&$4.74 \times 10^{-12}$ &$0.06$&  $0.28$ & $ 15\%$ & $ 100\%$\\
\\
U3  & Uncoupled & $6.00 \times 10^{-2}$&$12$ &$4.20 \times 10^{-11}$ &$0.06$&   $0.11$ & $0\%$ & $ 30\%$\\
C3  & Coupled & $6.00 \times 10^{-2}$& $12$ &$4.98 \times 10^{-11}$ &$0.06$&$0.17$& $15\%$ & $ 100\%$\\
\\
U4  & Uncoupled & $1.60 \times 10^{-1}$& $33$ &$2.57 \times 10^{-10}$ &$0.06$&  $0.05$& $0\%$ & $ 20\%$\\
C4  & Coupled & $1.60 \times 10^{-1}$& $33$ &$2.94 \times 10^{-10}$ &$0.05$&   $0.17$& $ 15\%$ & $ 100\%$\\ \hline
    \end{tabular}
    \caption{Summary of disc wind models, parameters, and results, in \textbf{cgs} units unless otherwise specified. The average disc luminosity $L_{\rm{avg}}$ is for coupled runs, averaged between 3000 and 3500 after variations drop to $\sim 1\%$. The mass flux $\dot{M}$ is measured as the outflow between 0 and 75$\degree$ at $10 r_{\star}$. We have listed the standard deviation $\sigma_X$ and maximum variations $\Delta X_{\rm{max}}$ for mass flux and disc luminosity over the time period $2200-3600 T_0$.}
    \label{tab:run_summary}
  \end{center}
\end{table*}

We consider two classes of models for line driven disc winds, where the disc intensity is computed from the accretion rate using (\ref{eq:intensity}). In \emph{uncoupled} models the accretion rate is assumed to be constant whereas \emph{coupled} models compute the disc intensity from the local, time-dependent accretion rate (\ref{eq:local_acc}). We study how the system behaves for four different average luminosities in both the uncoupled and coupled regimes. We control the luminosity, via the accretion rate, by varying the disc midplane density parameter, $\rho_0$, in (\ref{eq:density}). The parameters of the uncoupled runs were chosen to match the average accretion rate of coupled models at late times.  We summarize our list of models in Table \ref{tab:run_summary}. The uncoupled models are identical in setup to those in DP18b and used to benchmark the coupled models, which are the novel aspect of this work. 

Each run proceeds in two phases. First, the disc evolves for 2000 inner disc orbits $T_0$, with the radiation force turned off, to reach a quasi-steady accretion rate where  variations in luminosity are low ($\lesssim 15\%$). This ensures we can still explore variability while the equation (\ref{eq:intensity}), derived for a steady state disc can be used to approximate the resulting intensity. Over the next 100 $T_0$ the radiation force is turned on using a linear ramp up and we study the resulting disc-wind system for the following 1400  $T_0$. After this time, small variations in luminosity of $\sim 3\%$  persist as the disc accretion reaches a steady state . We have verified this for the fiducial run lasting over 9 000 $T_0$. 

\begin{figure}
  \includegraphics[width=0.48\textwidth]{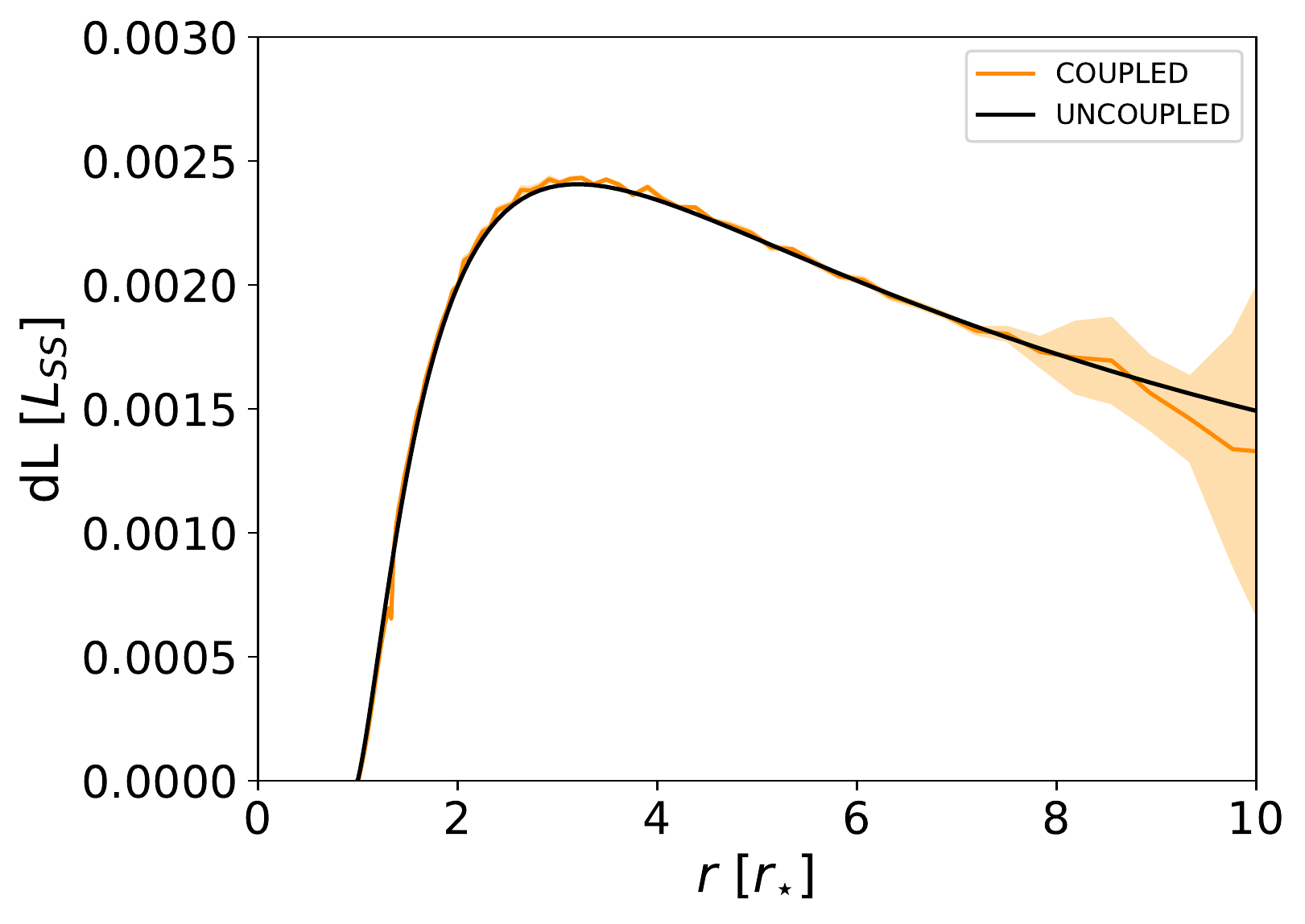}
  \caption{Time averaged luminosity of disc annuli $dL$ as a function of radius $r$ for $3500\leq t/T_0 \leq 4000$ for coupled (orange line) and uncoupled (black line) models, C2 and U2. The orange shading shows the standard deviation over this time interval. We see that variability at larger $r$ are persistent even into late times though the inner disc reaches a steady state.}
  \label{fig:luminosity}
\end{figure}

In Fig. \ref{fig:luminosity} we plot the time-averaged luminosity of an annulus (in units of the Shakura-Sunyaev disc luminosity, $L_{\rm{SS}}$), as a function of disc radius for uncoupled (black line) and coupled (orange line) models, U2 and C2, respectively. The shaded region shows the standard deviation in time of the luminosity for the coupled run. The variability in the outer regions of the disc persists into late times, albeit becomes very small. This convergence of the two classes of models at late times makes comparison of their resulting outflows easier.  

\subsection{Global Properties}
\begin{figure*}
    \centering
    \includegraphics[width=0.96\textwidth]{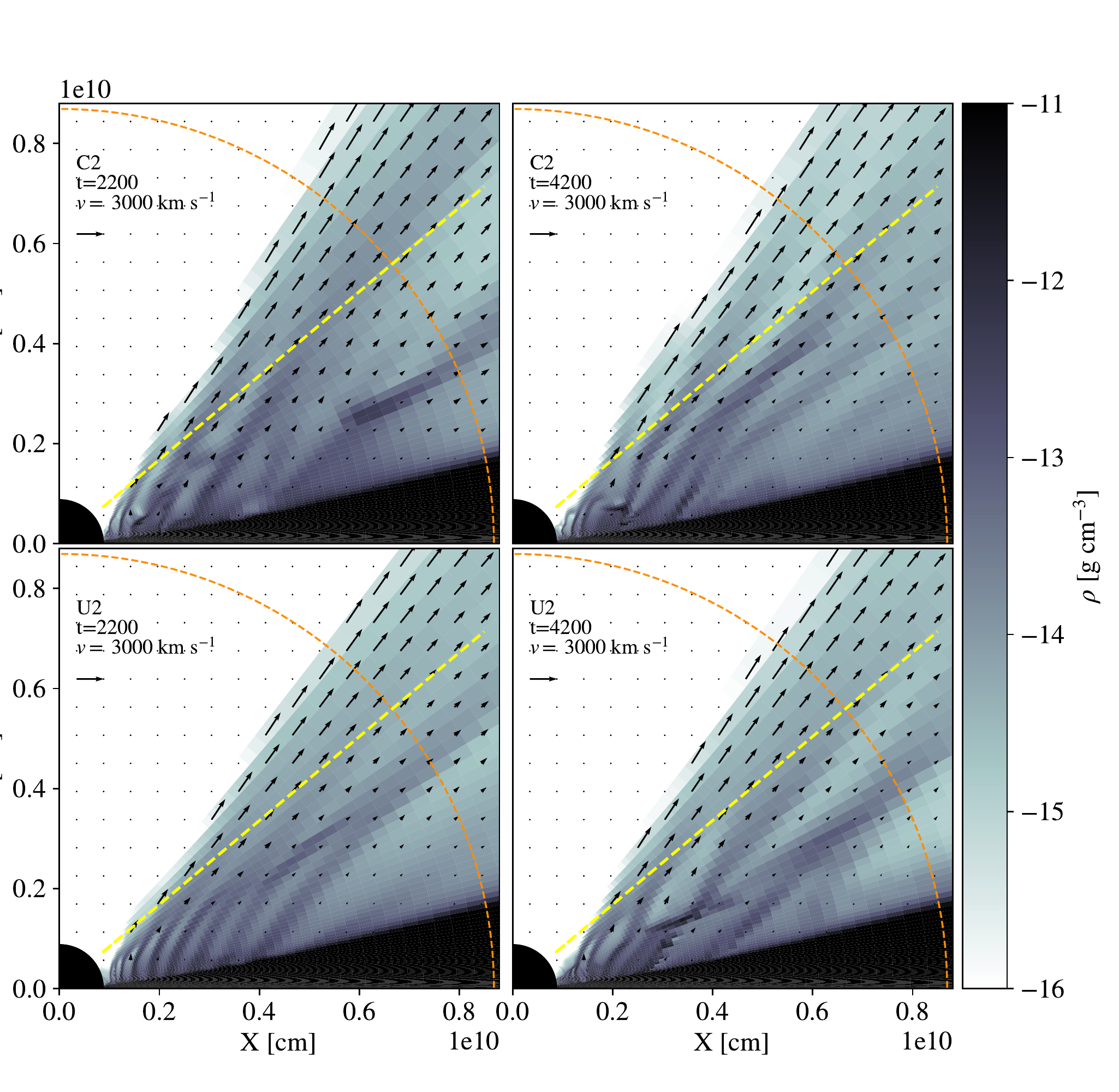}
    \caption{Yellow line at 50 degrees roughly indicating the separation between the fast and slow parts of the flow. The uncoupled run (U2) starts off ordered and stable (bottom left) but soon instability develops and the flow becomes turbulent (bottom right). In contrast, the coupled run is more turbulent at early times due to the varying radiation field (top left). The outflow is initially larger in the coupled run (C2, top left) than the uncoupled run due to periods of enhanced luminosity due to accretion spikes. This causes a ``luminosity peak'' in the flow (see Fig. \ref{fig:MdotLC1}, top right). At late times, the two runs are indistinguishable as luminosity variations drop below $3\%$ and any luminosity peaks are masked by the intrinsic variability of the wind.}
    \label{fig:Model2_density}
\end{figure*}

\begin{figure*}
  \centering
    \includegraphics[width=\textwidth]{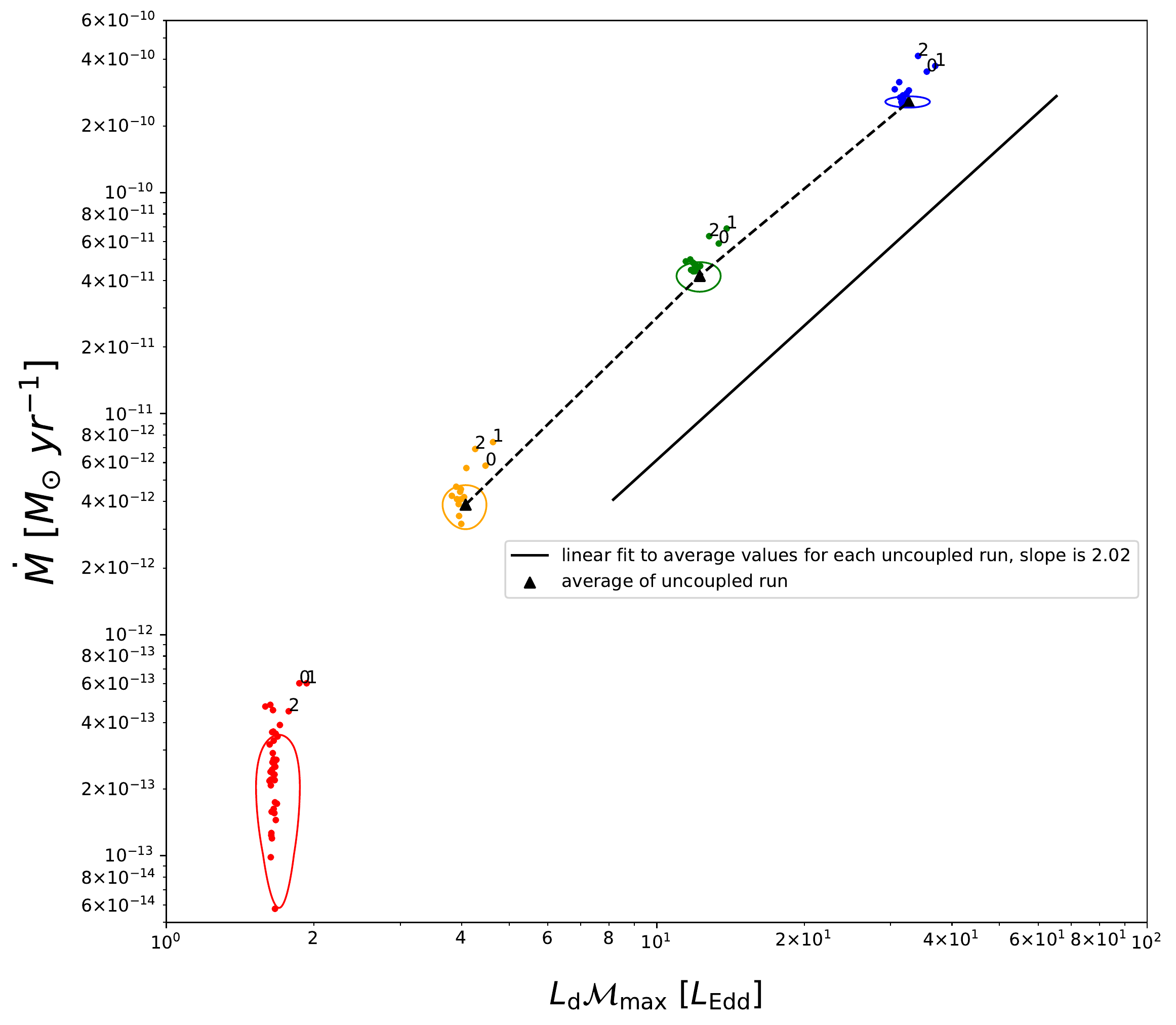}
    \caption{Outflow mass flux $\dot{M}$ and disc luminosity multiplied by the maximum of the line driving multiplier, $L_d {\cal{M}}\_\rm{max}$, for all disc wind models. Each colored point corresponds to the time-average over a 100 $T_0$ epoch from the coupled runs. Ellipses are defined by the max/min of the uncoupled runs. The disc evolution (captured in luminosity variation) is the same over all the models but there are multiple extreme points in the mass flux of the coupled runs (points outside the ellipses that have higher luminosity. At early times (epochs '0', '1', '2' are indicated on the figure) $\sim 15 \%$ variations in luminosity result in mass fluxes which deviate from the uncoupled models.}
  \label{fig:mdot_vs_L}
 \end{figure*}

We find global wind solutions in broad agreement with previous studies of line driven winds (see Fig. \ref{fig:Model2_density}). A conical outflow extending $\sim 45^{\circ}$ above the disc midplane, with the fastest parts of the flow at $45^{\circ}$ and the more radial flow being slower and denser. Further, the wind has small density structures, extending out from the disc, formed due to failed winds. These are winds that launch from the disc but fail to propagate to the outer boundary and fall back to the disc.

To understand the disc-wind system we search for correlations between the disc luminosity and global properties of the outflows. In Fig. \ref{fig:mdot_vs_L} we plot the wind mass flux as a function of the disc luminosity. Each color corresponds to a different wind model with each colored point the average for a 100 $T_0$ epoch for the coupled model. Ellipses indicate the one standard deviation in time contours for the corresponding uncoupled model. We see that the mass outflow scales approximately as $\dot{M} \propto L^{2}$, coming from the slope on Fig. \ref{fig:mdot_vs_L}. This is in agreement with previous studies for the region of phase space we explore, which is close to the turnover point where the dependence steepens. For very low luminosity the flow can even halt completely (Drew \& Proga 2000). The lowest luminosity run is not included in this fit since the relation turns over sharply near $L_d{\cal{M}}_{\rm{max}} \gtrsim 1$. 

\begin{figure*}
  \centering
    \includegraphics[width=0.48\textwidth]{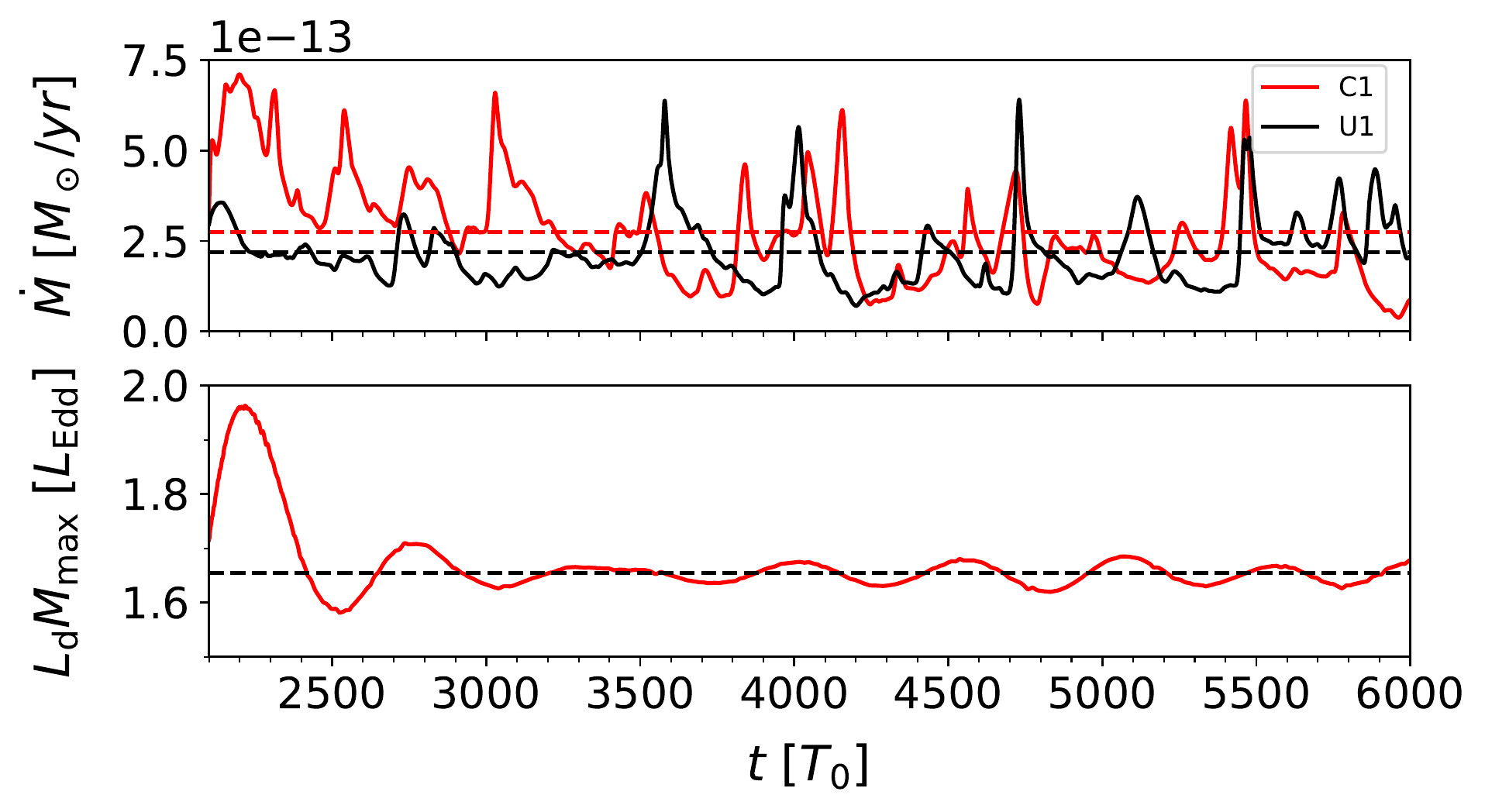}
        \includegraphics[width=0.48\textwidth]{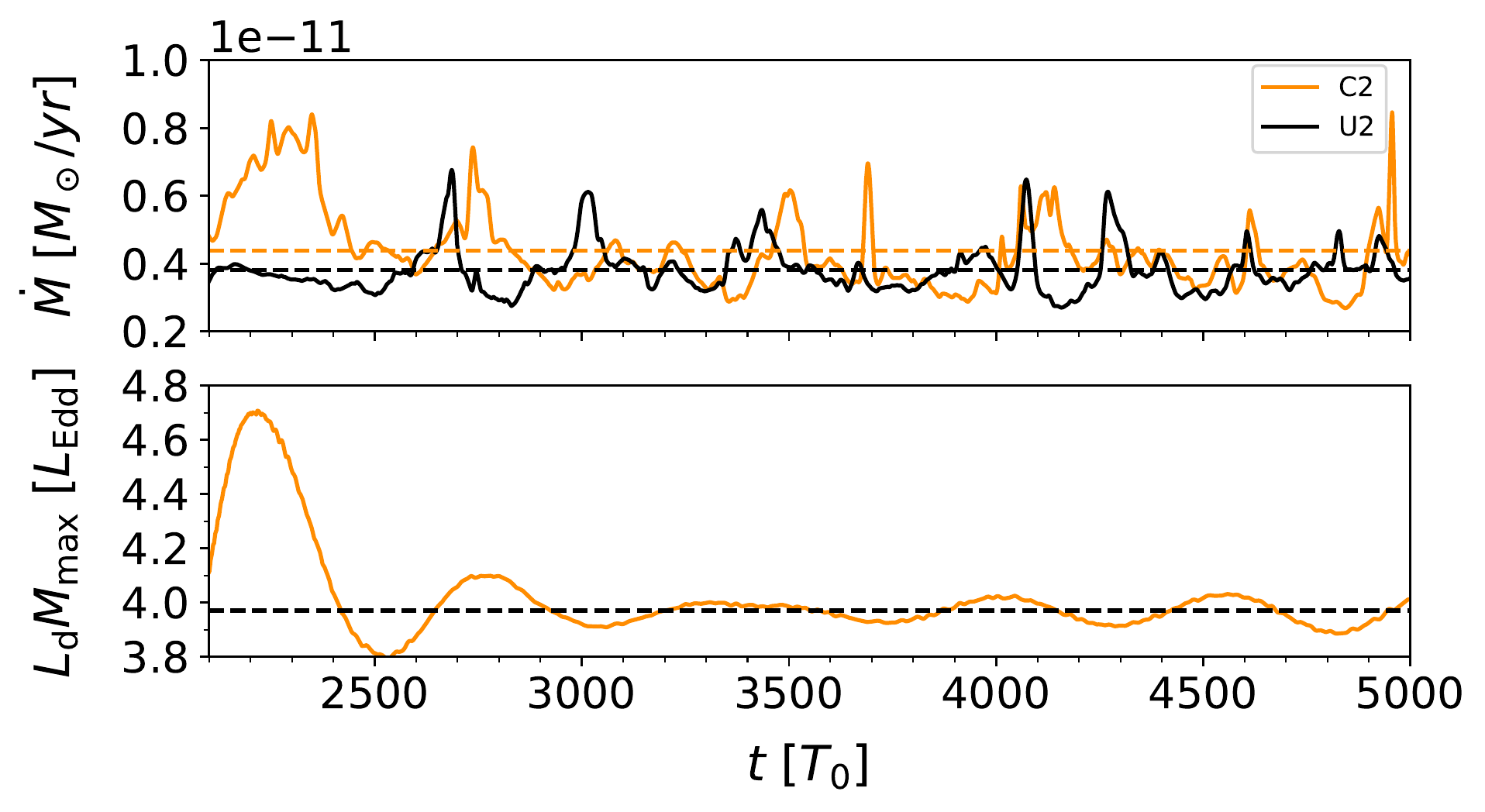}
         \includegraphics[width=0.48\textwidth]{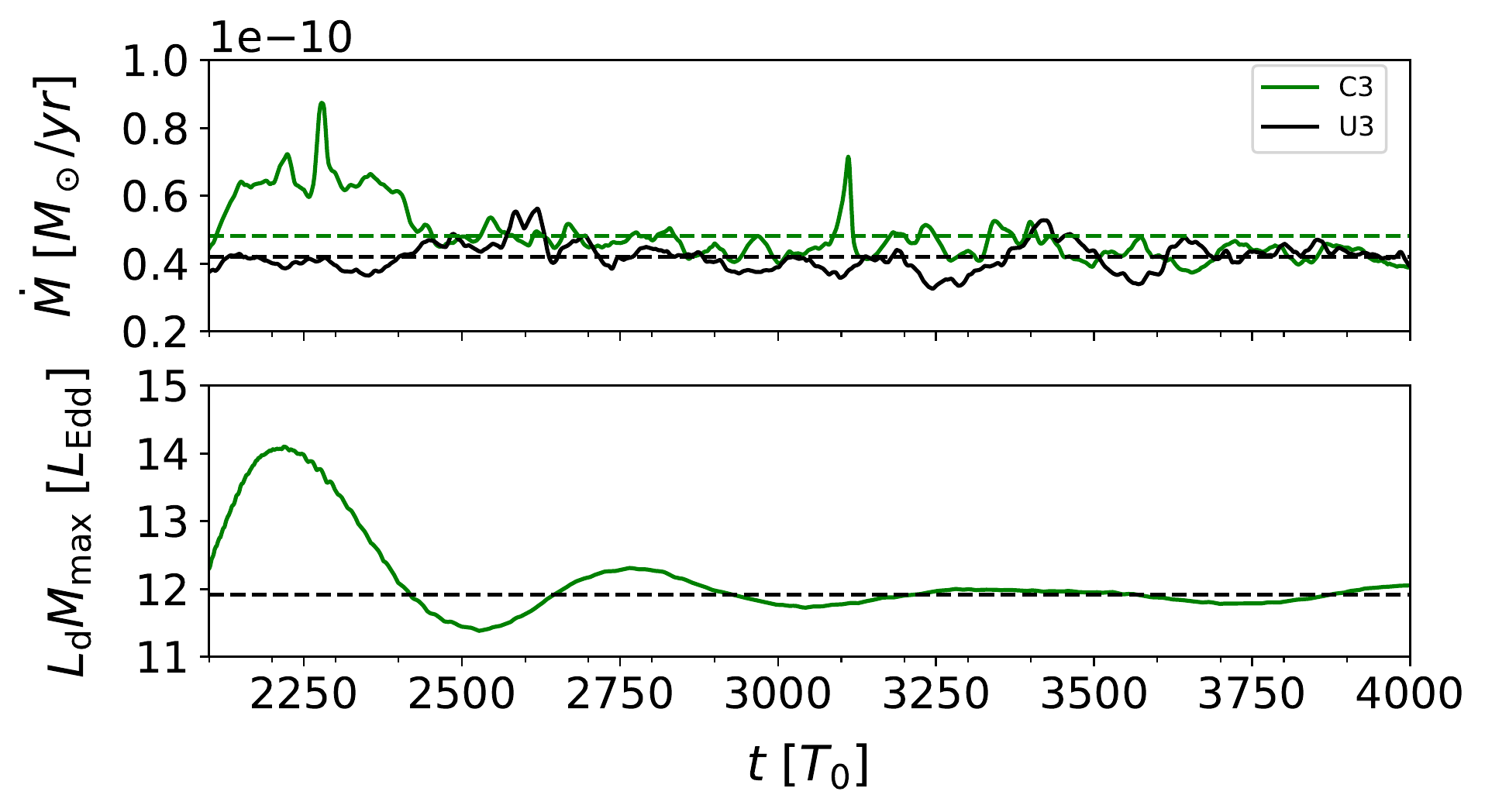}
         \includegraphics[width=0.48\textwidth]{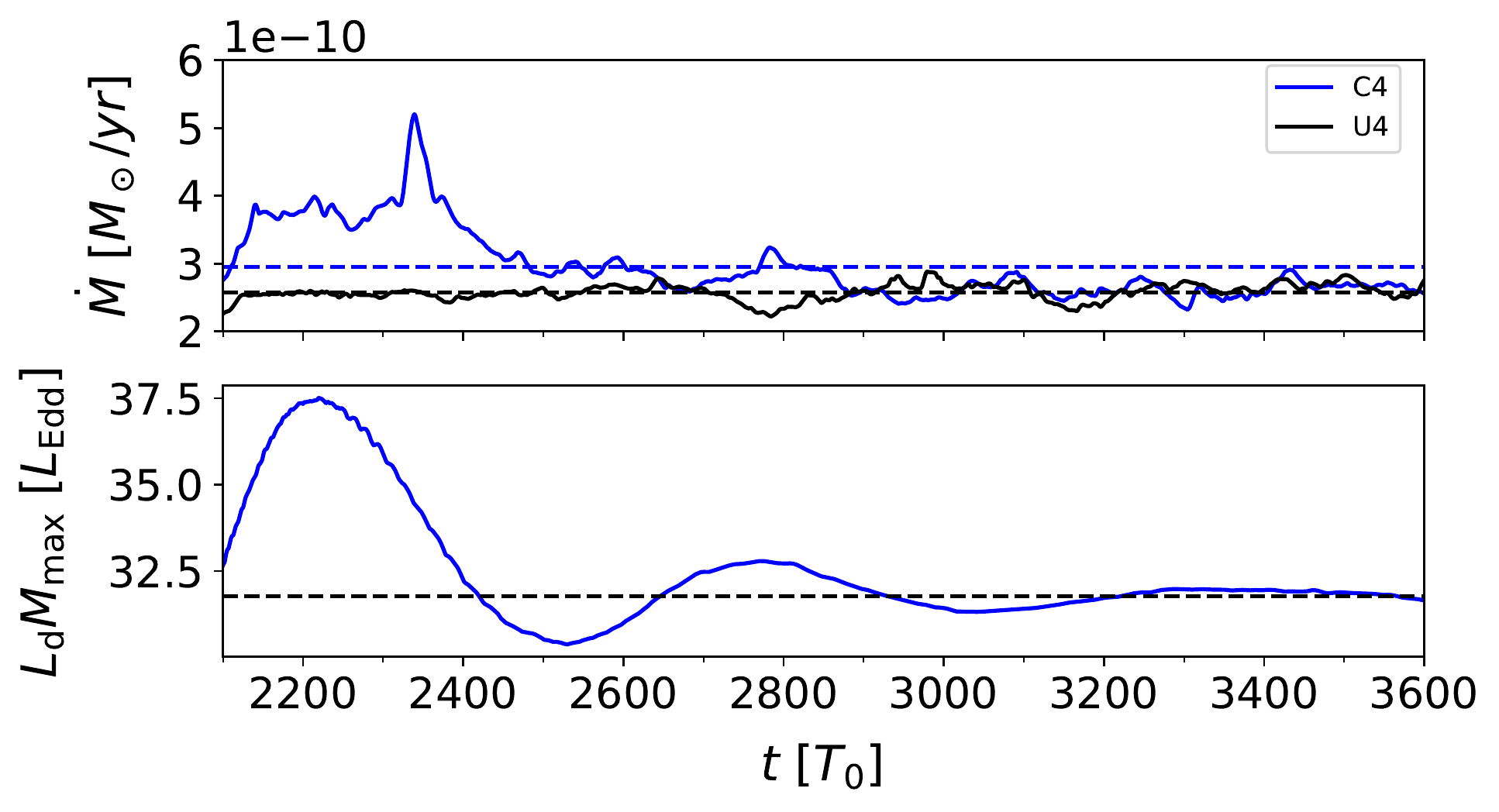}

    \caption{Outflow mass flux (upper panel) and disc luminosity (lower panel) for coupled (solid colored line) and uncoupled (dashed black line), with dashed lines indicating the temporal average. We show results for models 1 (top left), 2 (top right), 3 (bottom left) and 4 (bottom right). Uncoupled models have constant disc luminosity and the corresponding coupled models luminosity converges to this value at late times. Mass loss for coupled runs is higher initially due to the luminosity increase in the disc luminosity by $\sim15\%$. The luminosity peaks in the beginning are similar in height to the intrinsic peaks for low luminosity runs. Increases in mass outflow of $\sim3$ can be masked by intrinsic variability for Model 1. Note the second peak in the coupled run, C3, at $3100$. We track its origin to a clump that was most likely ejected by a local spike in the disc intensity, that did not visibly change the total disc luminosity. This peak occurs only in the coupled run and greatly exceeds the intrinsic variability at that late time. This illustrates how mass flux is affected by both local variations in disc intensity and total disc luminosity.}
    \label{fig:MdotLC1}
  \end{figure*}

The coupled runs exhibit greater variability, as evidenced by some epochs lying outside the uncoupled run contours. In particular, note that the epochs '0', '1', '2', which correspond to the first 300 $T_{0}$ of each run (after line driving is initiated), lie outside the contours. Those early times harbor the largest luminosity peak of each run, so their positioning outside of the contours makes sense (see Fig. \ref{fig:MdotLC1}). We will discuss this period in detail later in the paper. In relative terms, the higher luminosity runs are less inherently variable due to less small scale structure and a more smooth flow. In absolute terms however, the higher luminosity runs exhibit the largest mass flux and consequently the largest variability due to changes in luminosity.  This is also reflected in Table \ref{tab:run_summary}, where we have calculated the normalised standard deviation over the first 1400 $T_0$ after line driving is initiated. The normalised standard deviation, hereafter ``NSD'', for the uncoupled runs monotonically decreases with increasing luminosity, which shows that the higher luminosity runs are less inherently variable. All coupled runs have a higher NSD than their corresponding uncoupled run. The difference in NSD between uncoupled and coupled grows with increasing luminosity. For the highest luminosity models, the NSD differs from 0.05 to 0.17, while for the lowest luminosity run, the difference is negligible, 0.35 and 0.36 respectively. We identify two sources of variability in the mass flux. \emph{Intrinsic peaks} are due to the non-stationary nature of line driven winds and has been well documented in previous studies of line driven winds PSD98,99. \emph{Luminosity peaks} can be directly attributed to a spike in the disc luminosity. The NSD decreases with increasing luminosity due to a decrease in contributions from intrinsic peaks. For the low luminosity runs variability is due to both intrinsic and luminosity peaks, hence the coupled and uncoupled models have similar NSD. At higher luminosities, where outflows are more steady, outflow variability is dominated by luminosity peaks, hence the coupled models have $\sim3$ times higher NSD than the uncoupled models. As the flow is less steady for lower luminosity discs, local variations in disc intensity that are small relative to the total variability of disc intensity can alter the flow.  The lowest luminosity runs exhibit a larger average relative change and relative maximum change in mass outflow due to a luminosity change, as expected from the scaling in Fig. \ref{fig:mdot_vs_L} and reflected in the $\sigma_{\dot{M}}/\dot{M}_{\rm{avg}}$ and $\Delta \dot{M}_{\rm{max}}/\dot{M}_{\rm{avg}}$ values in Table \ref{tab:run_summary}. This turnover in the mass flux luminosity curve is particularly sharp near $L_d{\cal{M}}_{\rm{max}} \gtrsim 1$ where small changes in luminosity can result in winds failing to launch as the radiation force can barely overcome gravity.\newline
\indent In Fig. \ref{fig:MdotLC1} we plot the total disc luminosity and wind mass flux as a function of time for all models. The lower panel shows the total disc luminosity for the coupled (colored solid line) or uncoupled (black dashed line). The upper panel shows the time dependent outflow mass flux (solid line) and the time averaged value (dashed line) for the coupled (color line) or uncoupled (black line) model. We calibrated models by requiring that the late-time averaged luminosity of coupled and uncoupled models are equal.\\
\indent The coupled cases show clear evidence of correlations between the total disc luminosity and the mass flux. A $\sim 15\%$ variation in luminosity corresponds to a change in mass flux by a factor of up to $\sim 3$. These numbers depend on the total disc luminosity, with the lowest luminosity run showing the largest relative fluctuations. However, as discussed previously, these increases are more readily masked due to the inherent variability of the lower luminosity models. Therefore, the correlations are more apparent at higher luminosity where the uncoupled models have little intrinsic variability. The NSD difference between coupled and uncoupled models is a good proxy for how visible the correlations will be, with larger differences in NSD for stronger correlations between disc and outflow variability. Furthermore, we expect the  spatial location of this luminosity variation to play a role as well. Luminosity variations interior to a given fluid streamline are found to have a stronger effect than variations exterior to it (Dyda $\&$ Proga, 2018). In Section \ref{sec:lumpeaks} we examine how local, as opposed to global, variations in disc luminosity affect the outflow.\\
\indent As the disc tends to a steady state at late times variations in luminosity decrease to $\sim  3\%$ and the mass flux varies by $\sim  50\%$ in the C2 model. However, the uncoupled run has very similar variability, despite the radiation field being constant in time. At late times, variability correlated to disc luminosity becomes too small to reliably distinguish from the inherent variability seen in uncoupled models.\\
\indent We have seen that for lower luminosity runs, the inherent variability can mask mass flux variations due to luminosity increases (as is seen in  Fig. \ref{fig:MdotLC1} for C1 and C2). There are differences in the shape of the luminosity and intrinsic peaks. Intrinsic peaks are short and narrow. In contrast, luminosity peaks are wider, lasting on timescales of the fluctuations in luminosity.\\
\indent Finally, we also note that disc structure is largely unaffected by the wind. This is to be expected due to the disc being much more dense than the wind. Lower luminosity systems, on account of their lower mass winds, are expected to experience the smallest feedback on the disc. We observe $\lesssim 1\%$ change in disc luminosity across our suite of simulations, suggesting the feedback on the disc is negligible. While this applies in a straightforward manner to non-magnetic CVs, suggesting winds from these discs do not impact the discs themselves, AGN are more complicated due to their much higher luminosities, the need for ionization treatment and strong magnetic fields.

\subsection{Outflow Structure and Variability}
\label{sec:structure}
In this section, we characterize the local outflow structure and how it relates to global outflow properties. We discuss how the impact of luminosity variations on outflow structure result in different types of spikes in mass outflow and contrast them with spikes due to inherent variability. 

\begin{figure}
  \includegraphics[width=0.48\textwidth]{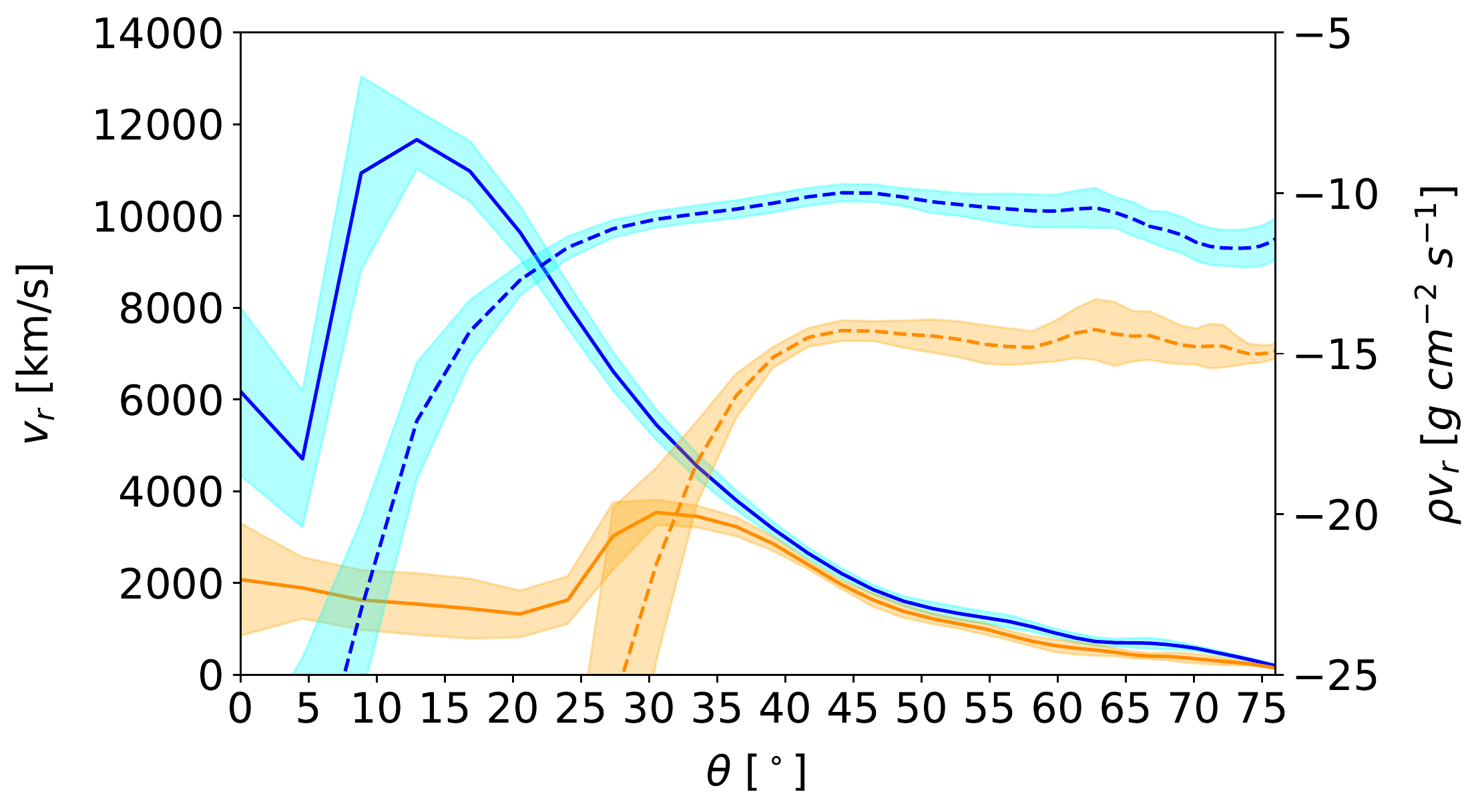}
  \caption{Time averaged momentum flux (dashed line) and radial velocity (solid line) as a function of polar angle at 10 $r_{\star}$ for $3000 \leq t \leq 3450$ for C2 (orange) and C4 (blue). The shading indicates standard deviation in time. In the higher luminosity model, the slow part of the flow is $\theta \gtrsim 35\degree$ , while for the lower luminosity $\theta \gtrsim 45\degree$.  }
  \label{fig:slowVSfast}
\end{figure}

In Fig. \ref{fig:slowVSfast} we plot the time averaged velocity (solid line) and momentum flux (dashed line) as a function of polar angle along $r = 10 r_{\star}$ for models C2, (colored lines) and U2 (black lines). The shading indicates one standard deviation of temporal variability during the epoch. The wind can be divided into ``fast'' and ``slow'' parts, the boundary between the two, we define as the angle after which the velocity has dropped below its minimum value along the rotation axis. 

The time dependent behaviour and structure of the flow are best seen in movies of the simulations.\footnote{Those can be seen at: \href{https://www.youtube.com/playlist?list=PLdaxJotZLlw3JYw58T2RqUkkhFP1E380Z}{https://www.youtube.com/playlist?
list=PLdaxJotZLlw3JYw58T2RqUkkhFP1E380Z}}
\subsubsection{Intrinsic peaks}
First, we discuss the intrinsic variability of the wind, present in both uncoupled and coupled runs. The boundary region between disc atmosphere and wind is particularly interesting. Close to the central object, there are vertical density structures, 'fingers', that can either be very ordered or very turbulent at different times (see Fig \ref{fig:Model2_density}). These structures, despite being close to the central object and more tightly gravitationally bound, are very important in determining the flow's variability. Often clumps form within them that are later either carried successfully to the outer boundary or fall back to the disc. In the case of the former, this leads to a peak in the mass loss rate. Clumps only exit in the ``slow'' part of the flow as inhomogeneities are smoothed out by the velocity shear in the ``fast'' stream. Hence, the slow stream is responsible for most of the variability in the mass flux.  Peaks in the momentum flux profile correspond to either the passage of a clump or a more global density fluctuation rather than to variations in the velocity field, which are insignificant. While the most variable parts in the velocity field are actually at the smallest angles, those parts of the flow are orders of magnitude less dense and contribute little to outflow variability. The flow at moderate values of $\theta$ is the least variable and then, once the above defined slow part starts, variability increases once again. Comparing momentum flux and velocity profiles as shown in Fig. \ref{fig:slowVSfast} for epochs both with and without peaks confirms that variations are due to changes in the slow parts of the flow. These intrinsic peaks occur for both the coupled and uncoupled runs and have been appreciated by the community since early disc wind models with static radiation fields. However, variability of disc intensity can further couple to produce additional variability, as we explain in the next section. 
\subsubsection{Luminosity Peaks}
\label{sec:lumpeaks}
Luminosity peaks refer to increases in the outflow rate correlated with increases in the total disc luminosity. In all coupled models, we observe that the initial peak in luminosity ($\sim 15\%$) is correlated with an increase in mass outflow. The flow is more turbulent close to the very inner regions of the disc (Fig \ref{fig:Model2_density}). This turbulence aids clump formation from early times in the coupled runs. In the uncoupled runs, the flow eventually becomes turbulent but less so as can be seen in the movies.  At later times, when the magnitude of variations in luminosity drops below 3\%, such luminosity peaks are less significant than that caused by intrinsic variability. In the higher luminosity runs, variations in luminosity are more strongly correlated with peaks in mass flux as intrinsic variability is weaker (see Fig \ref{fig:MdotLC1}). 

\begin{figure}
  \includegraphics[width=0.48\textwidth]{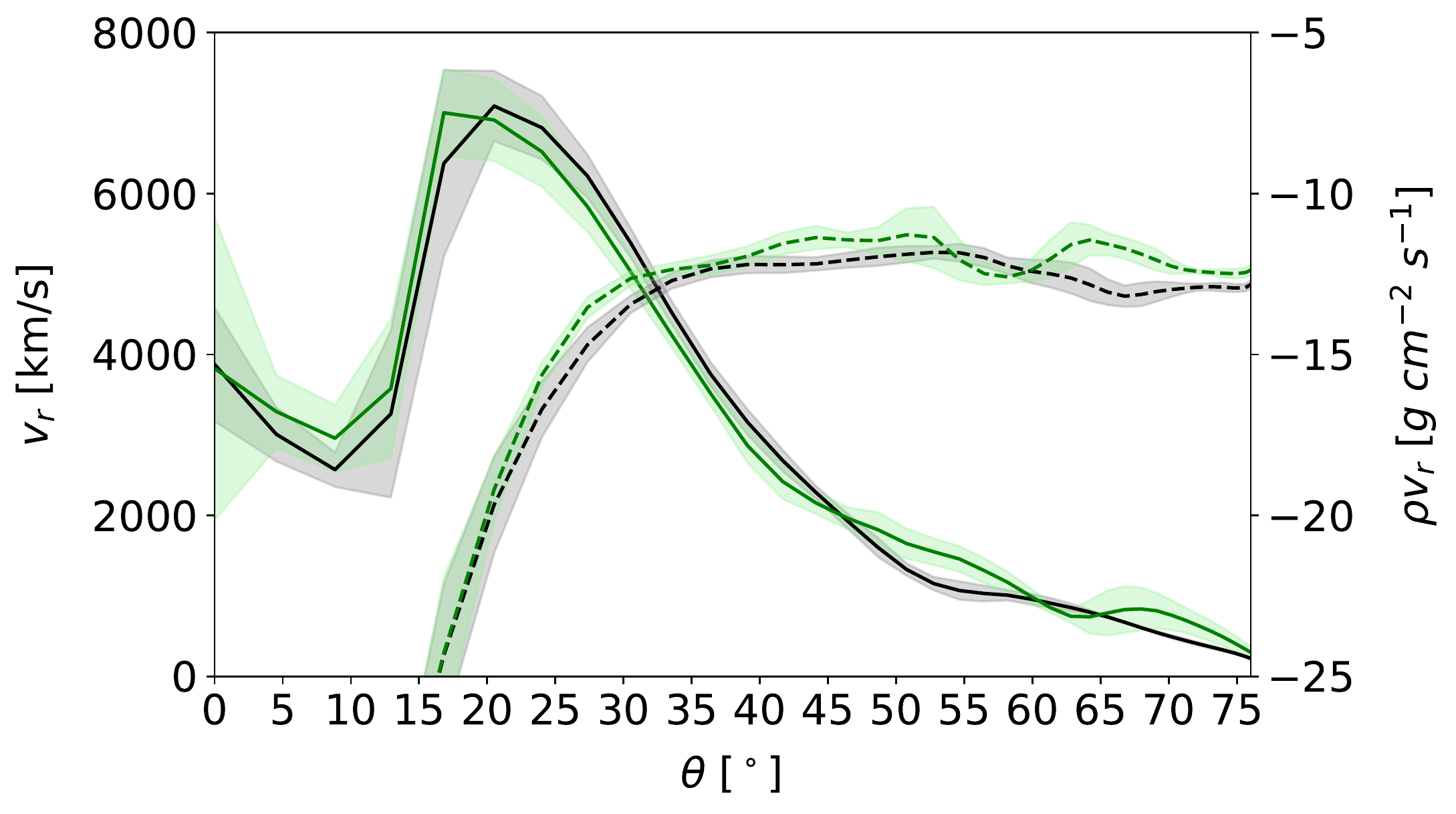}
  \caption{Time averaged momentum flux (dashed line) and radial velocity (solid line) as a function of polar angle at 10 $r_{\star}$ for $2200\leq t \leq 2300$ for C3 (green) and U3 (black). The shading indicates standard deviation in time. The coupled run experiences a luminosity peak during those times due to an increase in disc luminosity of $\sim 15\%$. The momentum flux differs by up to 1-2 orders of magnitude from the uncoupled run at certain angles ($45\degree$ and $65\degree$) while only the velocity varies slightly in comparison ($\sim 10-20\%$) and not at those angles. Hence, the dominant variability is in density fluctuations (through clumps), not velocity. }
  \label{fig:luminosity_peak}
\end{figure}

\begin{figure*}
  \centering
    \includegraphics[width=0.9\textwidth]{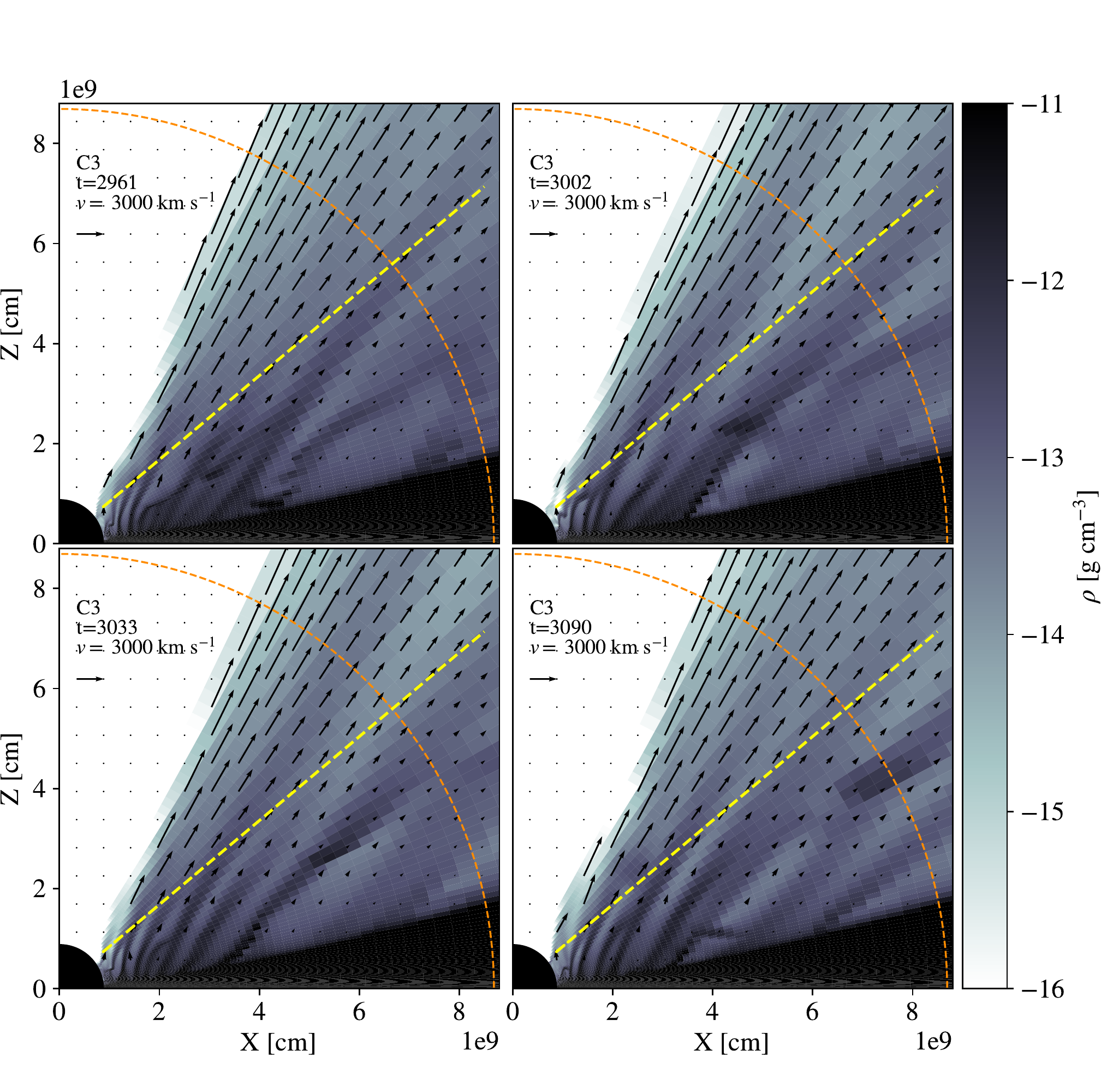}
    \caption{The propagation of clump aided by local luminosity increase for model C3. In the top left panel, the clump is forming at the base of the wind, close to the central object. In the top right panel, the clump is now visible just next to the yellow line, roughly halfway from the central object to the outer boundary. Shortly before that time, there was a $10\%$ increase in the luminosity directly below the clump that helped the clump to continue moving towards the outer boundary as opposed to falling back to the disc. The traces from this event are seen in the low density region that has formed below the clump in the top right panel. In the bottom left panel, the clump is seen propagating towards the outer boundary, with the low density region below it expanding. In the bottom right panel, the clump is finally seen crossing the outer boundary. This causes an increase in mass outflow at that time (around 3100) that is not seen in the uncoupled run (see Fig. \ref{fig:MdotLC1}, bottom left panel). This can be better seen in the movies of the simulation, found on our \href{https://www.youtube.com/playlist?list=PLdaxJotZLlw3JYw58T2RqUkkhFP1E380Z}{YouTube} playlist, in particualar, the C3/U3 movies. }
\label{fig:clump}
\end{figure*}

There are two ways in which an increase in luminosity causes an increase in mass outflow. If the total disc luminosity increases, there is a global increase in the mass outflow rate in both the fast and slow streams. (Fig. \ref{fig:luminosity_peak}). This happens on time scales of the dynamical time for gas moving in the fast stream. On the other hand, local increases in the disc intensity can affect the slow stream by accelerating clumps that would have otherwise failed to launch. Such clumps are responsible for the narrow structures superimposed on the broad luminosity peaks (see for example Fig. \ref{fig:MdotLC1}, lower right panel at $t = 2300$ and $t = 2800$). The temporal and spatial variability of the disc can combine to produce spikes in the outflow due to clumping. As an example, the mass flux spike in C3 at $t=3100$ (Fig. \ref{fig:MdotLC1}) was found to be produced by two clumps merging due to an increase in disc intensity directly below one of the clumps. The merging can also be seen in the C3 movie. This peak looks exactly like an intrinsic peak but its magnitude is much higher than expected for late times . Thus, while overall variations in total luminosity are  $\sim 3\%$ at late times, the increase in the \emph{local} intensity directly below the clump was $\sim 10\%$ and sufficient to alter the slow stream (see Fig. \ref{fig:clump}). This type of behaviour is of course impossible in the uncoupled runs. 

\section{Discussion and Conclusion}
\label{sec:discussion}
We studied  line driven disc winds with a time-dependent radiation field, computed self consistently from an accreting $\alpha$-disc. The central object was assumed to be much less luminous than the disc and to not contribute to the radiation field as expected for a non-magnetic CV. However, we expect some of our basic methods and results to be applicable to other line driven wind systems such as AGN. \\
\indent We see no significant feedback of the wind on the disc. For the Eddington parameters in this work, the ratio $\dot{M}_{\rm{out}} / \dot{M}_{\rm{acc}} \ll 1$, hence little angular momentum and energy is carried away by the wind, relative to the disc. We studied a different regime to Nomura et al. (2020), where a significant fraction of the accreted mass was lost to the wind and had to be accounted for. Furthermore, we do not study how the wind can impact the disc continuum as done by Nomura et al. (2020), which can be a source of feedback even for low luminosity systems and could be a direction for further work. Another interesting direction for further work in the case of CVs is to explore whether winds from the accretion disc can impact the accretion stream from the companion star.\\
\indent The intrinsic variability of the wind mass loss depends on disc luminosity. For the lowest luminosity systems, where $L_d {\cal{M}}_{\rm{max}} \gtrsim 1$, the intrinsic variability due to failed winds is comparable to luminosity peaks when total luminosity fluctuations are $\sim 15\%$ as we see at early times in our simulations. The main distinguishing feature between luminosity peaks and intrinsic peaks is their duration. Luminosity peaks occur on timescales of variability in the disc luminosity (with the exception of peaks caused by clumps aided by local disc variability), which can be much shorter than the viscous timescale of the disc. For a disc that is not in steady state, luminosity can vary on timescales from the shortest timescale (orbital) to the longest (viscous) (Pringle 1981). Indeed, we note luminosity variations on a shorter timescale than the viscous and much longer than orbital. Since our disc is almost in steady state, the timescale is closer to the former. Intrinsic peaks occur on time scales for clumps or voids to be ejected in the slow stream, that is to say a characteristic fluid crossing time. By contrast, when the luminosity is high, $L_d {\cal{M}}_{\rm{max}} \gg 1$, variability is dominated by luminosity peaks, provided the disc has not yet reached a steady state and luminosity still varies by $\gtrsim 10\%$. At late times, when the system approaches a steady state (luminosity variations $\lesssim 3\%$), correlations of mass loss with luminosity is lost in all runs as intrinsic variability dominates for all models.

Variability is dominated by the slow part of the flow. The variability is primarily due to density inhomogenieties, clumps and voids, which tend to be sheared away in the fast stream. The outflow's high degree of clumpiness and turbulence makes it sensitive to local variations in the disc intensity.  Often clumps or voids are formed in inner regions that propagate to the outer boundary and cause a spike in the slow part of the flow. Clumps can be aided by \emph{local disc intensity spikes} of $\sim10\%$, which do not appreciably alter the total disc luminosity, thereby creating a degeneracy between intrinsic and luminosity peaks. Thus, luminosity peaks in the mass flux can occur even though there is no apparent increase in the total disc luminosity (see model C3 at $t=3100$ in Fig. \ref{fig:MdotLC1}). Intrinsic variability of momentum flux happens through the slow parts of the flow primarily due to density fluctuations rather than velocity fluctuations. As can be seen in Fig. \ref{fig:luminosity_peak}, the intrinsic variations in flow velocity are much lower and similar for both high and low luminosity runs and uniform across the wind.  We further note that the $\sim 15\%$ changes in luminosity cause velocity increases in the low velocity flow. Those are more easily seen for systems with higher total system luminosity. We therefore expect variability due to small ($\lesssim 15\%$) variations in luminosity to primarily lead to absorption troughs becoming deeper as opposed to shifts in velocity space.  

Correlating wind activity with total system luminosity has thus far proven to be observationally challenging (Balman, Godon \& Sion 2014). Observations of V592 Cas (Kafka et al. 2009) have found no correlations between maximum velocity of absorption and the CV's brightness. Since the system is viewed at low inclination (\textit{low} inclination meaning that we are observing at \textit{low} $\theta$), only the highest velocity components would be observable as per the geometry of our model. Therefore, we expect that  $\sim15\%$  variations in luminosity to not have a significant effect on the observed maximum velocity of absorption as seen in Fig. \ref{fig:luminosity_peak}) , which is consistent with their observations of $\lesssim 10 \%$ luminosity variations. Furthermore, Kafka et al. (2009) propose that the non-axisymmetry of the flow could be due to a disc hotspot. In our simulations, we have seen that variations in disc intensity change the geometry of the flow. Hence, our work shows this is a plausible explanation, because of the evidence of phase modulation of the flow, but further study, perhaps using a persistent disc hot spot, as done by Cranmer \& Owocki (1996) in the context of stellar winds is needed. We note that they found a maximum wind velocity of $\sim$ 5000 $\rm{km}\   \rm{s}^{-1}$, consistent with our most luminous models. As discussed, the higher luminosity models show the most likelihood to have correlated disc and wind variability since the intrinsic variability of the wind is suppressed. Wind variability is expected to
be in the slow part of the flow, hence the ideal system would
be observed at high inclination.

For systems where we expect our line of sight to lie along the fast stream (low inclination systems), the best prospects are to observe transitions from a low to a high luminosity state or vice versa, the equivalent of a transition between two models in this work. In our simulations, larger variations in total driving luminosity can be extrapolated from the approximate scaling $\dot{M} \propto L^2$. For such large variations, one can expect more dramatic changes to the flow geometry. Depending on inclination, one might even expect to have periods where no wind is observed. In particular, if we observe a low inclination system that undergoes a dramatic drop in driving luminosity, the wind might become more equatorial and vice versa. For systems close to the $L_d {\cal{M}}_{\rm{max}} \lesssim 1$ threshold such state changes might also turn on/off the wind. 
In the system BZ Cam, estimates for inclination range from 12 to 40 degrees (Honeycut, Kafka \& Robertson 2013). This means that per our model, the disc would be observed at low $\theta$, much like V592 Cas. However, the luminosity variations in this system are large, meaning that we need to traverse from one model to another when the system transitions from a state of lower to a state of higher luminosity. The momentum flux at low inclination (corresponding to low $\theta$ in our model) would increase dramatically but mostly because of an increase in density (see Fig. \ref{fig:slowVSfast}), not velocity. This would still lead to an increase in line equivalent width of absorption. Indeed, such a correlation is seen in BZ Cam (Honeycut et al. 2013). The geometry and strength of the flow at lower $\theta$ is altered drastically between models (corresponding to large luminosity variations). At times of low luminosity, the flow might even stop completely at low $\theta$ as can be deduced from Fig. \ref{fig:slowVSfast} (traversing from model C4 to C2 would correspond to a large change in driving luminosity, which leads to flow between 20 and 30 degrees to cease almost completely). More studies of BZ Cam like the one done in Honeycut et al. (2013) are needed to clearly discern correlations between wind activity and luminosity.

A further challenge to observing these correlations is changes in ionization state. As proposed for BZ Cam by Greiner et al. (2001), changes in ionizing flux from the binary companion may explain changes in wind emission lines. Such a model predicts a periodic variability in ionization state set by the orbital timescale. Later observations by Honeycut et al. (2013) found variability on shorter timescales, disfavoring such a model. Further theoretical work should address this fundamental question of how variations in driving radiation (see for example Dyda et al. 2018d) can be distinguished from variations in ionizing flux from the outflow properties.

The ideal system to observe the effects of lower variability ($\lesssim 15\%$), corresponding to the variability within a model explored in this paper) is a high inclination system with sufficiently strong driving luminosity (so that the intrinsic variability is suppressed) and high enough accretion variability but a nearly constant ionizing flux. As can be seen on Fig. \ref{fig:luminosity_peak}, flow at lower $\theta$ is indistinguishable during a luminosity peak while at high $\theta$, both density and velocity increase during a luminosity peak (mainly density as can be judged from $\rho v_r$ changing by roughly an order of magnitude while velocity increases by less than $25\%$ on Fig. \ref{fig:luminosity_peak}). Therefore, looking for correlations between luminosity and line equivalent width of absorption in high inclination, high luminosity systems seems to be the most promising avenue for systems with low variations in luminosity ($\lesssim 15\%$). Even for such a system, correlations would be difficult to observe. One reason is that while mass outflow increases overall during a luminosity peak, that increase is not present in all parts of the flow (see Fig. \ref{fig:luminosity_peak}, e.g. flow at precisely 60 degrees is less during a luminosity peak even though flow around it increases significantly), so one might observe at an 'unlucky' inclination. A more pertinent reason is that intrinsic variability can mask peaks due to luminosity if driving luminosity is not sufficiently high. For example, Hartley et al. (2002) study two systems at inclinations of around 60 degrees, IX Vel and V3885 Sgr. The equivalent width of absorption for various lines is presented at three different times, where the continuum flux varies by $\lesssim 10\%$. No correlation is found between flux and wind, which could be due to the luminosity not being high enough so that the system is in the regime where variations due to luminosity dominate intrinsic variability. It is also possible that the inclination was particularly unlucky but that is unlikely. Since IX Vel is one of the brightest CV systems, one can reasonably ask the question whether a CV with 'sufficiently' high luminosity for these correlations to be visible even exists? A potential candidate system is ASAS J071404+7004.3 which is at a similar fortunate inclination of 62 degrees and seems to possess rapidly changing winds (Inight et al. 2022).

While we have chosen parameters more suited to CVs, some of our results are applicable to AGN. We expect higher inclination systems would be better for correlating small changes in luminosity to wind activity. AGN have much higher luminosities, so the intrinsic variability of the winds would be suppressed, making variations due to luminosity more visible. Low inclination systems (but not completely face on so that jet effects can be ignored) are better for correlating large variations in luminosity due to the changing geometry of the flow. However, in the regime of high luminosity variations, inclination is not as important because even systems viewed at high inclination (high $\theta$) are expected to have large fluctuations in density of flow due to changing luminosity (see Fig. \ref{fig:slowVSfast}). In the regime of small luminosity variations ($\lesssim 15\%$), AGN might be the only systems where correlations between luminosity and wind can be observed. We have also seen that local variations of the accretion rate in higher luminosity runs can 'conspire' to produce clumps and thus larger fluctuations in mass outflow than would be expected from the total luminosity variations (see Fig. \ref{fig:clump}). Further work, with overall much higher total luminosity and including ionization effects, is needed to understand line driven winds in AGN.

This paper motivates the need for more detailed self-consistent models of line driven winds from accretion discs to further understand how local variability can non-trivially influence mass outflow properties.

\section*{Acknowledgements}
SD and CSR acknowledge the UK Science and Technology Facilities Council (STFC) for support under the New Applicant grant ST/R000867/1 and the European Research Council (ERC) for support under the European Union's Horizon 2020 research and innovation programme (grant 834203). 

\section*{Data availability}
The data underlying this article will be shared on reasonable request to the corresponding authors.

\section*{References}
\begin{hangparas}{.25in}{1}

Balman S., Godon P., Sion E.M., 2014, ApJ, 794, 84

Castor, J.I., Abbott D.C., Klein R.I., 1975, ApJ, 195, 157

Clarke C., Carswell B., 2007, Principles of Astrophysical Fluid Dynamics. Cambridge University Press, Cambridge

Cranmer S. R., Owocki S. P., 1996, ApJ, 462, 469C

Drew J. E., Proga D., 2000, New Astron. Rev., 44, 21

Dyda S., Proga D., 2018, MNRAS, 475, 3786D

Dyda S., Proga D., 2018, MNRAS, 478, 5006D

Dyda S., Proga D., 2018, MNRAS, 481, 2745D

Dyda S., Proga D., 2018, MNRAS, 481, 5263D

Friend D.B., Abbott D.C., 1986, ApJ, 311, 701

Gayley K. G., 1995, ApJ, 454, 410

Greiner J., Tovmassian G., Orio G., Lehmann H., Chavushyan V., Rau A., Schwarz R., Casalegno R., Scholz R.D., 2001, A\&A, 376, 1031-1038

Hartley L. E., Drew J. E., Long K. S., MNRAS, 332, 127

Honeycut R. K., Kafka S., Robertson J.W., 2013, AJ, 145, 45

Inight K., Gänsicke B. T., Blondel D., Boyd D., Ashley R. P., Knigge C., Long K. S., Marsh T. R., McCleery J., Scaringi S., Steeghs D., Thorstensen J. R., Vanmunster T., Wheatley P. J., MNRAS, 510, 3605

Kafka S., Hoard D. W., Honeycutt R. K., Deliyannis C. P., 2009, AJ, 137, 197

Kurosawa R., Proga D., 2009, MNRAS, 397, 1791K
    
Liu C., Yuan F., Ostriker J. P., Gan Z., Yang X., 2013, MNRAS, 434, 1721L

Lucy L. B., Solomon P. M., 1970, ApJ, 159, 879

Mosallanezhad A., Yuan F., Ostriker J. P., Zeraatgari F. Z., Bu D., 2019, MNRAS, 490, 2567M

Murray N., Chiang J., Grossman S.A., Voit G.M., 1995, ApJ, 451, 498

Nomura M., Ohsuga K., Done C., 2020, MNRAS, 494, 3616N

Owocki S. P., Castor J. I., Rybicki G. B., 1988, ApJ, 335, 914

Parker E.N., 1958, PhFl, 1, 171P 

Pauldrach A., Puls J., Kudritzki R. P., 1986, A\&A, 164, 86

Pereyra N.A., 1997, PhD thesis, Univ. Maryland

Pereyra  N.A.,  Kallman,  T.R.,  Blondin,  J.M.,  1997,  ApJ, 477, 368

Pringle J.E., 1981, ARA\&A, 19, 137

Proga D., Stone J., Drew J.E., 1998, MNRAS, 295, 595 

Proga D., Stone J., Drew J.E., 1999, MNRAS, 310, 476

Shakura N.I., Sunyaev R.A., 1973, A\&A, 24, 337S

Stone J. M., Tomida K., White C. J., Felker K. G., 2020, ApJS, 249, 4S

Vitello P.A.J., Shlosman I., 1988, ApJ, 327, 680
\\
\\
\end{hangparas}

\appendix

\section*{Appendix A: Derivation of initial condition}
\renewcommand{\theequation}{A\arabic{equation}}
To improve the convergence time for the simulation, we introduced an appropriate initial condition with axisymmetry, $\rho=\rho(R,z)$ for our disc. Its derivation is as follows. 

Assuming hydrostatic equilibrium, in cylindrical coordinates $(R,z)$, a patch of gas must obey
\begin{equation}
\frac{1}{\rho}\dv{P}{z}=-\frac{GM}{R^2+z^2}\cos(\theta),
\end{equation}
where $\theta$ is the angle between the $\hat{R}$-axis and a radial vector pointing to from the central object to the patch of gas. If $z/R<<1$,
\begin{equation}
-\frac{GMz}{R^3(1+(z/R)^2)^{3/2}}\approx-\frac{GMz}{R^3}. 
\end{equation}
We can write this as,
\begin{equation}
\frac{1}{\rho}\dv{P}{z}\approx-\Omega_K^2 z,
\end{equation}
where $\Omega_K = \sqrt{\frac{GM}{R^3}}$, the Keplerian angular velocity.
The solution is an exponential in the $\hat{z}$-direction, assumming $z/R<<1$ and $P=c_s^2\rho$ with $c_s$ constant along $\hat{z}$ (vertically isothermal). Hence, we can write our density as\\
\begin{equation}
\rho = \rho (R) \rho_0' \exp(-z^2/2H^2),
\end{equation}
where $H=c_s/\Omega_K$ is the scale height. One can then relate the density to the surface density via integrating along $z$ from $-\infty$ to $+\infty$,
\begin{equation}
\Sigma = \sqrt{2\pi} H \rho(R).
\end{equation}
The surface density obeys a diffusion equation, assuming a Keplerian velocity profile. In the steady state, it can be shown (Clarke \& Carswell, 2007) that
\begin{equation}
\nu\Sigma = \frac{\dot{M}}{3\pi}\left(1-\sqrt{\frac{R_{\star}}{R}}\right),
\end{equation}\\
where $R_{\star}$ is the radius of the central object. This reveals the origin of the formula for $I(r_d)$ with $r_d\equiv R$ and $r_{\star}\equiv R_{\star}$. Recalling $\nu=\alpha c_s H$, and assuming isothermal equation of state (EoS) throughout (in $\hat{R}$ as well as $\hat{z}$), it also means that we can find $\rho(R)$, which we choose to write as
\begin{equation}
\rho(R) = \rho_0'' \left(\frac{R_{\star}}{R}\right)^3 \left(1-\sqrt{\frac{R_{\star}}{R}}\right),
\end{equation}
where we have absorbed all physical constants into $\rho_0''$. Now we can write $\rho(R,z)$ as

\begin{equation}
\rho(R,z) = \rho_0 \left(\frac{R_{\star}}{R}\right)^3 \left(1-\sqrt{\frac{R_{\star}}{R}}\right)\exp(-z^2/2H^2),
\end{equation}
where we $\rho_0=\rho_0'' \rho_0'$. Since we can choose $\rho_0'$, we can choose $\rho_0$. $\square$
\end{document}